\documentclass{elsart}
\usepackage{graphicx,harvard,amssymb}
\journal{New Astronomy}

\newcommand{\lsim}{\,\lower2truept\hbox{${<\atop\hbox{\raise4truept\hbox{$\sim$}}}$}\,}
\newcommand{\gsim}{\,\lower2truept\hbox{${>\atop\hbox{\raise4truept\hbox{$\sim$}}}$}\,}

\def\Ohat{\hat{\Omega}_b}
\newcommand{\eprime}{E^{\prime}}
\def\tsu27{\left(\frac{T_0}{2.7K}\right)}

\begin{document}

\begin{frontmatter}

\title{On the effect of cyclotron emission on the spectral 
distortions of the cosmic microwave background}

\author[Zizzo]{A.~Zizzo} and
\author[Zizzo]{C.~Burigana}

\vskip 0.5truecm

\address[Zizzo]{INAF-IASF, Sezione di Bologna,
Via Gobetti 101, I-40129, Bologna, Italy}
%\address[Burigana]{INAF-IASF, Sezione di Bologna,
%Via Gobetti 101, I-40129, Bologna, Italy}

\footnote{The address to which the proofs have to be sent is: \\
Carlo Burigana, INAF-IASF, Sezione di Bologna,
Via Gobetti 101, I-40129, Bologna, Italy\\
fax: +39-051-6398724\\
e-mail: burigana@bo.iasf.cnr.it}

\newpage
\begin{abstract}

We have investigated 
the role of the cyclotron emission associated to cosmic magnetic fields
on the evolution of CMB spectral distortions by considering the 
contributions by spontaneous and stimulated emission
and by absorption 
in the computation of 
the photon and energy injection rates.
These cyclotron emission rates
have been numerically compared 
with those of the relevant radiative processes
operating in the cosmic plasma, 
bremsstrahlung and double Compton scattering,
for realistic CMB distorted spectra
at early and late epochs.
For reasonable magnetic field strengths
we find that the
cyclotron emission contribution is much smaller than 
the bremsstrahlung and double Compton contributions, because of 
their different frequency locations 
and the high bremsstrahlung  
and double Compton efficiency to keep the long wavelength region of the 
CMB spectrum close to a blackbody 
(at electron temperature) during the formation of the 
spectral distortion.
Differently from previous analyses, we find that 
for a very 
large set of dissipation mechanisms
%, as in the case of 
%dissipation processes mediated 
%by energy exchanges between matter and radiation 
%or associated to photon injections at frequencies significantly
%different from the cyclotron emission frequency, 
%and for any realistic value of the cosmic magnetic field
the role of cyclotron emission 
in the evolution of CMB spectral distortions
is negligible 
and, in particular, it cannot re-establish a blackbody
spectrum after the generation of a realistic early distortion.
The constraints on the energy exchanges at various
cosmic times 
%set by the currently available data and 
%expected by future experiments
can be then derived, under quite general assumptions,
by considering only Compton
scattering, bremsstrahlung, and double Compton, other than,
obviously, the considered dissipation process.
Finally, upper limits to the CMB polarization degree induced
by cyclotron emission have been estimated.

\end{abstract}


\end{frontmatter}

\newpage

\section{Introduction}
\label{sec:intro}

The cosmic microwave background (CMB) spectrum emerges from the 
thermalization redshift,
$z_{therm}$,
with a shape very close to a Planckian one,
owing to the tight coupling between radiation and matter through
Compton scattering and photon production/absorption processes.
%radiative Compton and bremsstrahlung. 
Bremsstrahlung and double (or radiative) Compton 
were extremely efficient at early times
and able to re-establish a blackbody (BB) spectrum
from a perturbed one
on timescales much shorter than the expansion time,
$t_{exp}=a/(da/dt)$,
$a(t)$ being the cosmic scale factor and $t$ the time
(see, e.g., Danese and De Zotti (1977)).
Considering the effect of these processes combined to that of
Compton scattering,
the value of $z_{therm}$ ($\simeq 10^6 - 10^7$) \cite{buriganaetal91a}
depends on the baryon density parameter,
$\Omega_b$, and the Hubble constant, $H_0$, through the product
$\Ohat =\Omega_b [H_{0}/(50 {\rm Km} {\rm s}^{-1} {\rm Mpc}^{-1})]^2$.

On the other hand, physical processes occurring at redshifts $z < 
z_{therm}$ may lead imprints on the CMB spectrum.
Therefore, the CMB spectrum carries crucial informations on physical
processes occurring during early cosmic epochs
(see, e.g., Danese and Burigana (1993) and references therein)
and the comparison between models of CMB spectral distortions
and CMB absolute temperature measures can constrain the
physical parameters of the considered dissipation processes.

In the presence of a cosmic magnetic field, another  
photon production/absorption process,
the cyclotron emission,
operates in the cosmic plasma. 
The cyclotron emission (or synchrotron emission in the case
of relativistic particles) could be polarized 
\cite{rybickilightman}
and its degree of polarization
is an important indicator of the field's uniformity and structure
\cite{widrow}.

The contribution of the cyclotron emission to the evolution of
CMB spectrum depends on the amplitude of magnetic field
and on the electron density. Previous studies considered 
the cyclotron emission by including only the spontaneous emission term
\cite{puyepeter}
or by taking into account also the absorption and stimulated emission 
terms but assuming approximations
for CMB distorted spectra
that do not fully characterize the CMB spectral shapes 
realistically predicted in the presence of energy 
dissipation processes, in particular at long wavelengths 
where the cyclotron emission occurs \cite{afshordi}.

In this work we derived the contribution of the cyclotron emission
to the evolution of the CMB photon occupation number, $\eta$, 
as a further term in the Kompaneets equation \cite{kompaneets}
by exploiting the method described by Afshordi (2002)
%\cite{afshordi} 
to take into account the cyclotron spontaneous emission,
absorption and stimulated emission terms 
and generalizing it to be able to exhaustively treat various 
reasonable choices for the photon occupation number. 
We then apply this result to realistic assumptions for the 
CMB distorted spectra in order to provide robust estimates
of the global photon production rate as a function of the relevant
parameters and discuss the
role of the cyclotron emission in the thermalization 
and evolution of CMB spectral distortions.

In Sect.~\ref{sec:campi-magnetici} we briefly report on the 
main observational and theoretical aspects of cosmic magnetic fields
relevant for the present work. The contribution of the cyclotron
emission associated to cosmic magnetic fields to the evolution
of the CMB spectrum is derived in Sect.~\ref{sec:effect}, where
the cyclotron frequency is compared with the other 
characteristic frequencies relevant in this context.
In Sect.~\ref{sec:results} we compare 
the production rates
of photon number and energy densities 
from the cyclotron emission with those from bremsstrahlung
and double Compton for two different realistic cases
of distorted spectra, Bose-Einstein like and Comptonization like 
distortions. The typical values of the cosmic
magnetic field for which the  
cyclotron emission contribution to thermalization process could 
be comparable to the contribution from bremsstrahlung
and double Compton are derived.
Finally, we discuss our results and draw the main conclusions
in Sect.~\ref{conclusion}.

\section{Magnetic fields in cosmology}
\label{sec:campi-magnetici}

Magnetic fields pervade the universe at different scales
(see, e.g., Vall\'ee (2004), Gaensler et al. (2004)
%\cite{gaensleretal04} 
for recent reviews), from 
the scale of planet and stars to the scales of galaxies and clusters 
of galaxies and of the whole universe, possibly affecting
the cosmogonic process (see, e.g., Subramanian and Barrow (1998),
Rees (2000)). 
%\cite{subramanianbarrow,rees}). 
In this section we briefly report on the main 
observational results on cosmic magnetic fields and on physical 
models for their generation in the early 
universe and their evolution (see, e.g., 
Kronberg (1994), Grasso and Rubinstein (2001),
Carilli and Taylor (2002), Widrow (2002)
%\cite{kronberg,grassorubinstein,carilli,widrow}
for reviews).

\subsection{Observations of magnetic fields}
\label{sect:osservazioni_di_B}
The main observational traces of Galactic and extragalactic magnetic 
fields are the Zeeman splitting of spectral lines, 
the intensity and polarization
of the synchrotron emission from 
free relativistic electrons, 
and the Faraday rotation measurements (RM)
of the polarized electromagnetic radiation passing through a ionized 
medium.\\%
The Zeeman splitting, though direct, is typically too small to be usefully 
used for observations outside the Galaxy.\\%
The RM of the radiation emitted by a source at a redshift $z_s$ is given 
by \cite{grassorubinstein,widrow}
\begin{equation}\label{eq:Faraday_rotation}
RM(z_s) \simeq 8.1\times10^5\int_0^{z_s}
{n_e \over {\rm cm}^{-3}} {B_{\parallel} \over \mu{\rm G}}%
(z)(1+z)^{-2}dl(z)\frac{\mbox{rad}}{\mbox{m}^2},
\end{equation}
where $B_{\parallel}(z)$ is the magnetic field strength along the line of 
sight and
\begin{equation}\label{eq:lineofsight}
dl(z) \simeq {c \over H_0}(1+z)^{-1}
[\Omega_m(1+z)^3 + (1-\Omega_m-\Omega_\Lambda)(1+z)^2+\Omega_\Lambda]^{-1/2}
dz \, ;
\end{equation}
here $\Omega_m$ and $\Omega_\Lambda$ are the matter and 
cosmological constant (or dark energy) density
parameter, and 
$n_e$ is the electron number density along the considered line 
of sight.

The interstellar magnetic field in the Milky Way has been %
determined using several methods which allowed to obtain valuable %
information about the amplitude and spatial structure of the field. %
The average field strength is found to be  $\sim 3-4\mbox{ }\mu$G. %
Such a strength corresponds to an approximate energy equipartition %
between the magnetic field, the cosmic rays confined in the Galaxy, and %
the small-scale turbulent motion \cite{kronberg} 
\begin{equation}\label{eq:Equipartition}
\rho_m=\frac{B^2}{8\pi}\approx\rho_t\approx\rho_{CR} \, .
\end{equation}
The field keeps its orientation on scales of the order of few Kpc, %
comparable with the Galactic size, and two reversals have been %
observed between the Galactic arms, suggesting that the Galaxy %
field morphology may be symmetrical. %
Magnetic fields of similar intensity have been observed in a number %
of other spiral galaxies.\\ %
Observations on a large number of Abell clusters \cite{kimetal}, %
some of which have a measured X-ray emission, give %
valuable informations about magnetic fields in clusters of galaxies. %
The magnetic field strength in the intracluster medium (ICM) %
is well described by the phenomenological equation 
\begin{equation}\label{eq:B_ICM}
B_{ICM} \simeq 2\left(\frac{L}{10\mbox{kpc}}\right)^{-1/2}
[H_{0}/(50 {\rm Km} {\rm s}^{-1} {\rm Mpc}^{-1})]^{-1} 
\mbox{$\mu$G} \, ,
\end{equation}
where $L$ is the reversal field length.
Typical values of $L$ are $\approx 10-100$~Kpc corresponding
to field amplitudes of $1-10\mbox{ }\mu$G.\\% 
High resolution RMs towards very distant quasars %
have allowed to probe magnetic fields in the distant past. %
The measurements are consistent with an average field %
strength of $0.4-4\mbox{ }\mu$G on a coherence length of %
$\sim 15$Kpc, comparable with a typical galaxy size.\\%
The RMs towards distant quasars are also used to constrain the intensity %
of magnetic field in the intergalactic medium (IGM). %
Assuming an aligned cosmic magnetic field, 
the RMs of distant quasars imply $B_{IGM}\lsim 10^{-11}$~G
(for a simple cosmological model with 
$\Omega_{tot}=\Omega_m=1,\mbox{ }\Omega_\Lambda=0$, and
$H_{0}=75 {\rm Km} {\rm s}^{-1} {\rm Mpc}^{-1}$).
Unfortunately the largest reversal scale is at most $\sim 1$ Mpc. %
By adopting this scale and applying Eq.~(\ref{eq:Faraday_rotation}) %
the limits on $B_{IGM}$ are less stringent, $B_{IGM}\lsim 10^{-9}$~G
at present time.
\subsection{Generation of early magnetic fields}
\label{sect:genesisofB}

Quantum field theory provides a large numbers of %
possible physical mechanisms that may generate magnetic fields in the early  %
universe.\\%
Magnetogenesis typically requires a non-thermal equilibrium and %
a macroscopic parity violation. %
This condition could have been satisfied during the phase transitions 
(PT) in the early stages of the universe.\\%
Some authors have shown that magnetogenesis is possible during %
phase transitions in the quantum chromodynamics (QCD) era. %
During the %
quark-hadron phase transition a baryon-excess build %
up in front of the bubble wall, just as a consequence of %
the difference of the baryon masses in the quark and hadron phases
\cite{chengolinto}. % 
In this scenario magnetic fields are generated by the peculiar %
motion of the electric dipoles which arises from %
convective transfer of the latent heat released by the %
expanding bubble walls. %
The field strength at the QCDPT time has been estimated by the %
authors to be $B_{QCD}\simeq 10^8\mbox{ G}$ on a maximal coherence %
length $l_{coh}\simeq H_{QCD}^{-1}$.\\ % 
The magnetic field 
on scales $L\gg\mathit{l}$ can be estimated by performing a proper volume %
average of the fields produced by a large number of magnetic %
dipoles of size $\mathit{l}$ randomly oriented in space
\cite{hogan}.
Such an average gives %
\begin{equation}
\label{eq:Hogan}
B_L=B_l\left(\frac{l}{L}\right)^{3/2}.
\end{equation}
Using  Eq.~(\ref{eq:Hogan}) the strength of the magnetic field 
on the galactic 
length scale at the present time is found to be
$B(\mbox{kpc})\simeq 10^{-20}$G.

Some of the ingredients which may give rise to magnetogenesis %
may also be found at the electroweak phase transitions (EWPT).\\ %
Strong magnetic fields %
can be generated by a first order EWPT via dynamo mechanism
\cite{baymetal}. %
The authors estimated the average magnetic field strength %
at the present time 
$B(R\sim 10^9\mbox{ AU})\sim 10^{-17}-10^{-20}\mbox{ G}$.%

Finally, the existence of a magnetic field at decoupling 
may induce a Faraday rotation in the CMB polarization signal.
For example, for a field strength of $\simeq 10^{-9}$~G 
Kosowsky and Loeb (1996) derived a rotation of $\simeq 1^\circ$ 
at $\simeq 30$~GHz possibly observable by future CMB
polarization experiments. 

\subsection{Evolution of cosmic magnetic fields}

The time evolution of a magnetic field in a conducting medium is 
described by the equation
\cite{jackson}
\begin{equation}\label{eq:diffusionediB}
\frac{\partial{\bf B}}{\partial t}=\nabla\times({\bf v}\times{\bf B})+
\frac{c^2}{4\pi\sigma}\nabla^2{\bf B} \, ,
\end{equation}
where $\sigma$~ is the  electric conductivity. %
Neglecting fluid velocity $\mathbf{v}$, this equation reduces to the 
diffusion equation which implies that an initial magnetic 
configuration will decay away in a time
\begin{equation}\label{eq:diffusiontime}
\tau_{diff}(L)=\frac{4\pi\sigma L^2}{c^2} \, ,
\end{equation}
where $L$ is the characteristic length scale of the spatial variation of %
$\mathbf{B}$. In a cosmological framework, this means that a magnetic %
configuration with coherence length $L_0$ will survive until the %
present time $t_0$ only if $\tau(L_0)>t_0$. In our convention $L_0$ %
corresponds to the present time length scale determined by %
the Hubble law
\begin{equation}\label{eq:coherencelength}
L_0=L(t_i)\frac{a(t_0)}{a(t_i)} \, ,
\end{equation}
where $a(t)$ is the cosmic scale factor and $L(t_i)$ is the length scale %
at the time of the formation of the magnetic configuration. %
As shown by Eq.~(\ref{eq:diffusiontime}) the relevant quantity %
controlling $\tau_{diff}$ is the electric conductivity of the medium. %
This quantity changes in time depending on the varying population of %
the available charge carriers and on their kinetic energy. %
%However, since most of the universe evolution takes place %
%in a matter dominated regime, 
Assuming that all charge carriers %
are non-relativistic, the estimate of magnetic diffusion length is simple. %
For simplicity, we consider only one charge carrier type, the 
electrons, 
with charge ${\it e}$, 
number density 
${n_e}$, mass $m_e$, and velocity ${\bf v}$. Comparing the Ohm law 
${\bf J}=\sigma{\bf E}$ %
with the current density definition ${\bf J}=n_ee{\bf v}$, and using 
the expression ${\bf v}\sim e{\bf E} \Delta \tau/m_e$ for the mean drift 
velocity 
in the presence of the electric field $\mathbf{E}$, $\Delta \tau$ being 
the average time between collisions of the considered charge carrier, 
for the electron %
conductivity we have:
\begin{equation}\label{eq:conduttivitaelettrica}
\sigma=\frac{n_ee^2 \Delta \tau}{m_e}.
\end{equation}
For the evolution of the electron number density 
$n_e$ we can use here~\footnote{In this work, we are focussing
on the pre-recombination era when matter were highly ionized.} 
the usual formula 
\begin{equation}\label{eq:neevolution}
n_e\simeq (7/8)n_b \simeq 2.45 \times 10^{-6}\,\Ohat (1+z)^3 \mbox{ 
cm}^{-3} \, ,
\end{equation}
where a primordial helium 
abundance of $\simeq 25\%$ by mass
has been assumed here for numerical estimates
(in this hypothesis 
$n_H\!=\!0.75n_b\mbox{, and }n_{He}\!=\!(1/16)n_b
\mbox{, and }n_b=2.8 \times 10^{-6}\,\Ohat (1+z)^3 \mbox{ cm}^{-3}$).
Since electron resistivity is dominated by Thomson scattering 
off cosmic background photons then 
$\Delta \tau \simeq 1/n_{\gamma}\sigma_T c$, 
where 
$\sigma_T=8\pi/3(e^2/mc^2)^2$ is the Thomson cross section, and 
therefore  Eq.~(\ref{eq:conduttivitaelettrica}) gives
\begin{equation}\label{eq:conduttivitaelettrica1}
\sigma=\frac{n_e e^2}{m_e \sigma_T n_{\gamma} c} \, ,
\end{equation}
where
$n_{\gamma} \simeq n_P \simeq 
4 \times 10^2 (T_0/2.7{\rm K})^3 (1+z)^3$~cm$^{-3}$ 
is the photon number density in the blackbody limit, 
$aT_0$ being the present CMB 
energy density
($T_0 \simeq (2.725 \pm 0.002) {\rm K}$; Mather et al. (1999)).
%\cite{mather99}). 
The high conductivity of the cosmic medium has a relevant 
consequence for the evolution of magnetic fields.
The magnetic flux through any loop moving with fluid is a 
conserved quantity in the limit $\sigma\rightarrow\infty$. 
In fact, the diffusion equation (\ref{eq:diffusionediB}) 
after few vector algebra operations implies 
\begin{eqnarray}\label{eq:flussodiB}
\frac{d\Phi_S({\bf B})}{dt}=\int_{S(t)}\frac{\partial{\bf B}}
{\partial t}-\nabla\times({\bf v}\times{\bf B})d{\bf S}=
-\frac{c^2}{4\pi\sigma}\int_{S(t)}\nabla\times\nabla\times
{\bf B}\cdot d{\bf S} \, ,
\end{eqnarray}
where ${\bf S}$ is any surface delimited by the loop. %
On scale where diffusion can be neglected the field is said to be %
{\it frozen-in}, in the sense that lines of force move together with %
the fluid. Assuming that the universe expands isotropically, 
magnetic flux conservation implies 
\begin{equation}\label{eq:evoluzionediB}
{\bf B}(t)={\bf B}(t_0)\left(\frac{a(t_0)}{a(t)}\right)^2={\bf B}_0(1+z)^2,
\end{equation}
where ${\bf B}_0$ is the present time magnetic field. %

\section{Effect of the cyclotron emission associated to cosmic magnetic 
fields on the evolution of CMB spectral distortions}
\label{sec:effect}

The evolution of the CMB photon occupation number, $\eta(\nu,t)$, 
at redshifts %
$z\lsim10^{6}-10^{7}$, is well described by the Kompaneets equation %
\cite{kompaneets}. %
We can write formally this equation as
\begin{equation}\label{eq:Kompaneetsformal}
\frac{\partial\eta}{\partial t}=\left(\frac{\partial\eta}
{\partial t}\right)_{\Lambda}+\left(\frac{\partial\eta}
{\partial t}\right)_{\Gamma}=\sum\Lambda_i+\sum\Gamma_i \, ,
\end{equation}
where $\Lambda_i$ take into account processes that do not change the 
photon number and $\Gamma_i$ take into account 
photon production/absorption processes.\\%
During the cosmic epochs of interest here, before recombination, the 
relevant %
processes are Compton scattering, bremsstrahlung (BR) and double (or 
radiative) 
Compton (DC). %
We want to study the role of magnetic fields on the evolution 
of the CMB photon occupation number. 
The effect of a magnetic field in a ionized plasma is to speed up all 
the present charged particles. We consider here only the electrons.\\
We shall show in this section how to calculate the contribution 
$(\partial\eta/\partial t)_{CE}$ in Eq.~(\ref{eq:Kompaneetsformal}) 
due to the cyclotron emission (CE) of 
electrons 
accelerated by a cosmic magnetic field in the primeval plasma. 
Then we compare this contribution with that of bremsstrahlung 
and double Compton in the case of early and late CMB spectral 
distortions by using 
appropriate analytical formulas for the description of the 
photon occupation number $\eta$.
\subsection{Cyclotron emission}
\label{sec:emissionediciclotrone}

Following the approach presented  by Afshordi (2002)
%\cite{afshordi} 
the rate of energy loss via 
cyclotron emission by non-relativistic 
electrons moving in a magnetic field ${\mathbf B}$ 
can be obtained classically 
\cite{jackson}
\begin{equation}\label{eq:energyloss}
\frac{d \mathcal{E}}{d t}=\frac{2}{3}\frac{e^2\omega_c^2 \langle{v_{\perp}}^2 
\rangle n_e}{c^3}=\frac{2}{3}\frac{e^4B^2 \langle{v_{\perp}}^2 
\rangle n_e}{m_e^2c^5} ;
\end{equation}
here $\omega_c=2\pi\nu_c=eB/m_ec$ is the cyclotron frequency, 
$E_c=h\nu_c=\hbar\omega_c$ is the energy of the emitted photon, $v_{\perp}$ 
is the component of the electron velocity normal to the 
magnetic field
direction.
% and $n_e$ the electron number density. 
In the non relativistic limit, almost all the emitted photons have the 
frequency $\omega_c$ and thus the rate of photon production per unit 
volume, $\psi$, can be obtained using Eq.~(\ref{eq:energyloss})
\begin{equation}\label{eq:numberdensityofcyclotron}
\psi=\frac{d{\mathcal E}/dt}{\hbar\omega_c}=\frac{2}{3}\frac{n_e e^3 B 
\langle{v_{\perp}}^2\rangle}{\hbar m_e c^4} \, .
\end{equation}
By assuming a Maxwellian distribution for the electrons we have
\begin{equation} 
\langle{v_{\perp}}^2\rangle=\frac{2}{3}\langle v^2\rangle=
\frac{2k_B T_e}{m_e} \, ,
\end{equation}
where $T_e$ is the temperature of the electron gas. %
Eq.~(\ref{eq:numberdensityofcyclotron}) 
then becomes 
\begin{equation}\label{eq:psi}
\psi=\frac{4}{3}\frac{n_e e^3 B k_B T_e}{\hbar {m_e}^2 c^4} \, .
\end{equation}
The presence of photons in the environment enhances the photon production 
through stimulated emission. Also photons can be absorbed by %
the rotating electrons. %
These processes can be expressed via \cite{spitzer}
\begin{equation}\label{eq:ratediciclotrone1}
\!\!\!\!\!\!\!\!\!
\left(\frac{\partial\eta(E^{\prime})}{\partial t}\right)_{CE}\!\!\!
=\left[\sum_{\{E\}}{\mathcal A}(1+\eta(E^{\prime}))\eta_e(E+E_c)-
{\mathcal B}\eta(E^{\prime})\eta_e(E)\right]\delta(E^{\prime}-E_c),
\end{equation}
where $\eta_e$ is the electron distribution,  ${\mathcal A}\:\mbox{ e }\:
{\mathcal B}$ are the Einstein coefficients, $E^{\prime}$ is the photon 
energy, 
 $E+E_c$ is the electrons energy and $\delta(E^{\prime}-E_c)$ is the Dirac 
$\delta$ function. The sum is over the energy states of the electrons 
(Landau levels).\\
The first two terms in the second member of 
Eq.~(\ref{eq:ratediciclotrone1}) describe
the spontaneous and stimulated emission, the third term
is the contribution due to the absorption of photons: a photon of energy
$E^{\prime}=E_C$ can be emitted by an electron which undergoes a decrease 
of energy from $E+E_c$ to $E$, or may be absorbed by an electron of 
energy $E$ 
(which clearly raises its energy to $E+E_c$). 
The coefficient ${\mathcal B}$ can be obtained considering that for %
a Planck distribution $\partial\eta/\partial t\equiv 0$; therefore
\begin{equation}\label{eq:coefficienteB}
\sum_{\{E\}}e^{(E^{\prime}/k_BT_e)}{\mathcal A}\eta_e(E+E_c)
=\sum_{\{E\}}{\mathcal B}\eta_e(E) \, . 
%\nonumber
\end{equation}
By assuming a Maxwellian distribution for the electrons, in the non %
relativistic limit we can write
\begin{eqnarray*}
\sum_{\{E\}}\eta_e(E)=\frac{1}{h^3}\int 8\pi m_eE^2e^{-E/k_BT_e}dE=n_e
\end{eqnarray*}
from which one obtains
\begin{eqnarray*}
\int \frac {8\pi m_e}{h^3} E^2\eta_e(E+E_c)dE=n_e\;e^{-E_c/k_BT_e} \, .
\end{eqnarray*}
From these equations together with Eq.~(\ref{eq:coefficienteB}) we have
\begin{eqnarray*}
{\mathcal B}={\mathcal A}\;\;e^{(E-E_c)/k_BT_e} \, .
\end{eqnarray*}
Substituting this expression in Eq.~(\ref{eq:ratediciclotrone1}) is
straightforward to obtain
\begin{equation}\label{eq:ratediciclotrone2}
\left(\frac{\partial\eta(E^{\prime})}{\partial t}\right)_{CE}={\mathcal A}\,
n_ee^{-E_c/k_BT_e}[1+\eta(E^{\prime})
-e^{E^{\prime}/k_BT_e}\eta(E^{\prime})]\delta(E^{\prime}-E_c) \, .
\end{equation}
We can integrate Eq.~(\ref{eq:ratediciclotrone2}) over the 
phase space to have the total photon injection rate
\begin{equation}\label{eq:rategenerica}
\frac{d n}{d t}=\frac{2}{h^3}\int 4\pi \frac{{\eprime}^2}
{c^3}\left(\frac{\partial\eta(\eprime)}{\partial t}\right)_{CE}d\eprime 
\, ;
\end{equation}
here the factor $2$ takes into account the possible polarization of the 
photon and the coefficient ${\mathcal A}$ in 
Eq.~(\ref{eq:ratediciclotrone2})
is the photon production rate 
for zero photon occupation number, i.e. the same as $\psi$ in 
Eq.~(\ref{eq:psi});
inserting 
the expression 
for $\partial\eta/\partial t$ 
given by Eq.~(\ref{eq:ratediciclotrone2}) 
in Eq.~(\ref{eq:rategenerica}) 
one finds 
%
%\begin{eqnarray*}
%\left(\frac{\partial n}{\partial t}\right)_{CE}&=&
%\frac{8\pi{\mathcal A}n_e}{c^3h^3}
%e^{-E_c/k_BT_e}\int {\eprime}^2 \delta(\eprime-E_c)d\eprime\\
%&=&\frac{8\pi n_e{\mathcal A} E_c^2}{h^3c^3}=
%\psi
%\end{eqnarray*}
%
%from which
%
\begin{equation}\label{eq:coefficienteA}
{\mathcal A}=\frac{4\,\pi^2\,e\,c\,k_B\,T_e}{3\,B}\,e^{E_c/k_BT_e} \, .
\end{equation}
By introducing the dimensionless frequency $x_e=\eprime/k_BT_e$, 
from 
Eqs.~(\ref{eq:coefficienteA}) and (\ref{eq:ratediciclotrone2}) 
we have
\begin{equation}
\label{eq:ratesdicy}
\left(\frac{\partial \eta}{\partial t}\right)_{CE}=K_{CE}(z)
\,[1-\eta\,(x_e)\,(e^{x_e}-1)]\,\delta(x_e-x_{e,CE}) \, ,
\end{equation}
where
\begin{equation}\label{eq:Kcy}
K_{CE}(z)=\frac{4\pi^2\,e\,c}{3 B(z)}\,n_e=4.64\times10^{-4}\,
\Ohat\cdot B^{-1}(z)(1+z)^3\mbox{ s}^{-1} \, .
\end{equation}

The magnetic field will be assumed to scale 
as in Eq.~(\ref{eq:evoluzionediB}).\\
Now we can write the complete Kompaneets 
equation including cylotron emission 
\begin{eqnarray}\label{eq:kompaneets+ciclotrone}
\frac{\partial\eta}{\partial t}&=&\frac{1}{\phi}\frac{1}{t_C}
\frac{1}{x^2}\frac{\partial}{\partial x}\left[x^4\left[\phi
\frac{\partial\eta}{\partial x}+\eta(1+\eta)\right]\right]\\
\nonumber\\
&+&\left[K_{BR}\frac{g_{BR}}{x_e^3}e^{-x_e}+K_{DC}\frac{g_{DC}}{x_e^3}+K_{CE}
\delta(x_e-x_{e,CE})\right]\left[1-\eta(e^{x_e}-1)\right] \, , \nonumber
\end{eqnarray}
where the coefficients $K(z)$ and the Gaunt factors,
$g_{BR}$ and $g_{DC}$, for %
bremsstrahlung \cite{karzaslatter,rybickilightman}
and double Compton \cite{gould84}
are given in 
Burigana et al. (1991) and Burigana et al. (1995),
%\cite{buriganaetal91a,buriganaetal95},
$t_C=mc^2/[kT_e(n_e \sigma_T c)]$ 
is the timescale for the achievement of kinetic equilibrium between
radiation and matter,
$\phi=T_e/T_r$ 
where $T_r=T_0(1+z)$ is the CMB temperature, and
$x=\phi x_e$.\\ %
At this point a little consideration on the cyclotron term is necessary. %
In a realistic framework cyclotron emission should not be %
a line emission, as considered here, but it should be a continuum emission 
peaked around the characteristic frequency $\nu_{CE}$. In addition, 
fluctuations in the plasma and in the magnetic field imply analogous
fluctuations in the cyclotron emission frequency. 
Therefore a better approximation should involve 
a replacement of 
the $\delta$ function approximation 
with a proper (physically motivated) smooth 
function in all numerical estimates. 
On the other hand, 
as it will be evident from 
the results presented in the next sections,
the simple $\delta$ function approximation
does not significantly limit the validity of all our main conclusions
that rely on the typical magnitude of the cyclotron emission
frequency as function of redshift and not on its detailed 
value.

\subsection{Numerical comparison between characteristic frequencies}
\label{sec:confrontitrafrequenze}

In this section we numerically compare the %
characteristic frequencies appearing in the %
``classical'' (i.e. in the presence of C, DC, and BR) solutions of the 
Kompaneets equation with that of cyclotron emission. %
We can write the dimensionless frequency of cyclotron emission as follow:
\begin{eqnarray}\label{eq:xcy}
x_{e,CE}={h\nu_c \over kT_e} \simeq 
4.97\times 10^{-5}\cdot \phi^{-1}\tsu27 ^{-1}B_0(1+z),
\end{eqnarray}
where $B_0$ is the present value of the magnetic field.
Let us define the frequency $x_{e,c}$ as the solution 
of the equation $t_{abs}=t_C$ 
\cite{sunyaevzeldovich,illarionovsunyaev}, %
where $t_{abs}$ is defined by \cite{buriganaetal91a,buriganaetal95}
\begin{eqnarray}\nonumber
1/t_{abs}&=&\left[E_{BR}(x_e,z)e^{-x_e}+E_{DC}(x_e,z)\right](e^{x_e}-1)\\
\nonumber\\
&=&\left[\frac{K_{BR}(z)g_{BR}(x_e)e^{-x_e}}{x_e^3}+
\frac{K_{DC}(z)g_{DC}(x_e)}{x_e^3}\right]
(e^{x_e}-1) \, .
\label{eq:tabs}
\end{eqnarray}
Let us define also 
the frequency $x_{e,abs}$ according to the relation
\cite{danesedezotti}:
\begin{equation}
\label{eq:defxabs}
%y_{abs}(x_e)=\tau_0-\tau=
%\int_1^{1+z}\frac{t_{exp}}{t_{abs}}\frac{d(1+z)}{1+z}=1.
y_{abs}(x_e, t, t_h)=\tau (t) -\tau (t_h)=
\int_{1+z}^{1+z_h}\frac{t_{exp}}{t_{abs}}\frac{d(1+z)}{1+z}=1 \, 
\end{equation}

when $y_{abs}$ is evaluated at the current time $t_0$ ($z=0$);
here $t_h$ ($z_h$) is the time (redshift) at which a given
distortion occurred.
%

%%%%%%%%%%%%%%%%%%%%%%FIGURA X_C%%%%%%%%%%%%%%%%%%%%%%%%%%%%%%%%%%%%%%%%%%%%%%
%\begin{figure}[!htbp]
\begin{figure}[!th]
\vskip 1cm
\begin{center}
%\special{psfile=xc.ps  hscale=85 vscale=65 hoffset=-50 voffset=-360}
\includegraphics{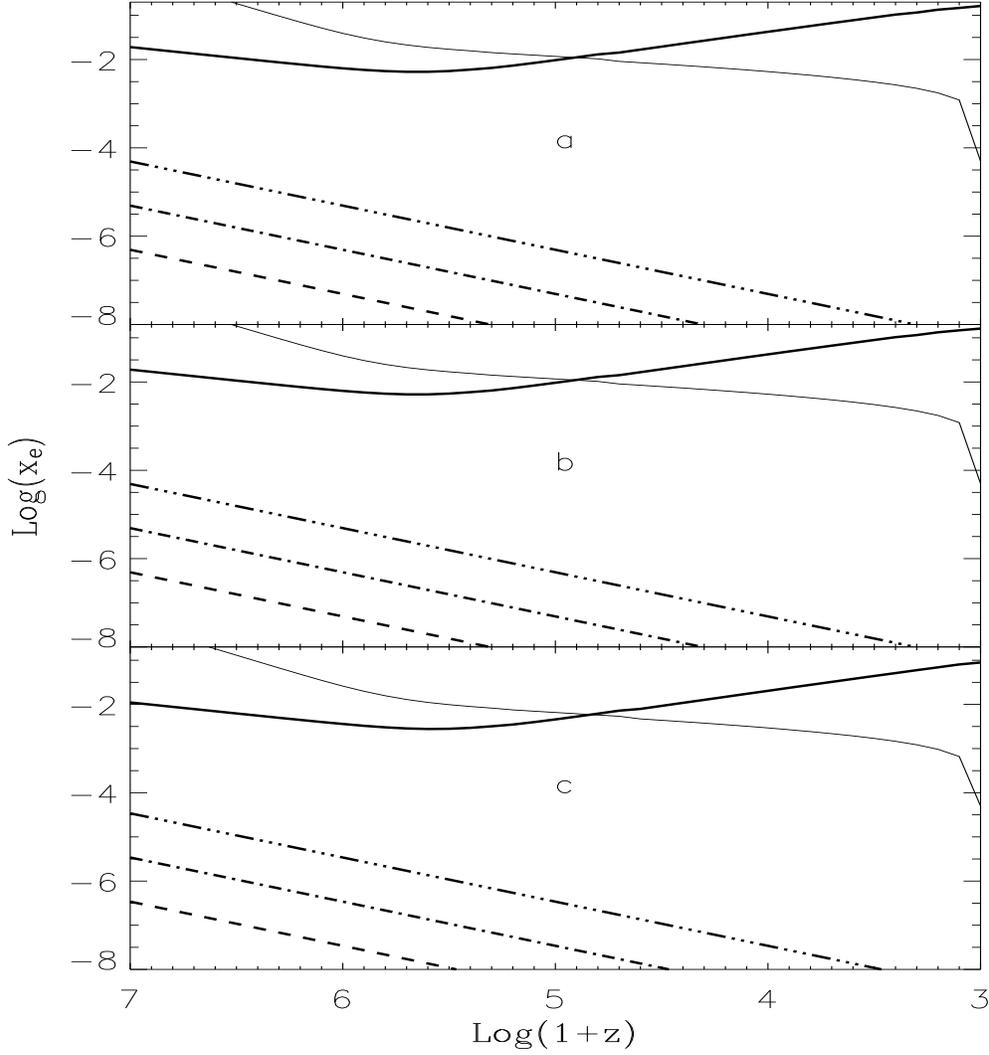}
\vspace{12.5 cm}
\linespread{1} 
\caption{Comparison between the characteristic frequencies $x_{e,c}$ 
(thick solid line), $x_{e,abs}$ (thin solid line) and $x_{e,CE}$, for 
different values of the magnetic field $B_0$:
$10^{-9}$G (thick dashed line), $10^{-8}$G  
(thick dott-dashed line), $10^{-7}$G (thick three dots and dash line).
In the three panels different values 
of the fractional energy 
$\Delta\varepsilon/\varepsilon_i$, 
injected in the CMB radiation field are assumed. 
We consider the case of early (Bose-Einstein like) distorted spectra
with $\Delta\varepsilon/\varepsilon_i$ ($\simeq \mu_0/1.4$ for 
$\mu_0 \ll 1$) values corresponding to
the following values of the chemical potential $\mu_0$ (see also 
Sect.~\ref{sec:earlydistortion}): $10^{-4}$ 
(panel {\it a}), $10^{-2}$ (panel {\it b}), $1$ (panel {\it c}). 
The cosmological parameters relevant in this context
have been assumed in agreement with WMAP (see Tab.~3 
of Bennett et al. (2003));
%\cite{bennett}); 
nevertheless the dependence on the detailed choice of the assumed cosmological 
parameters is here not particularly critical.
%$\Omega_b=0.044,\;H_0=71\; 
%z_{eq}=10^4\mbox{ and }z_{rec}=10^3$.
}
\label{fig:xc}
\end{center}  
\end{figure}
\linespread{1.3}
%%%%%%%%%%%%%%%%%%%%%%%%%%%%%%%%%%%%%%%%%%%%%%%%%%%%%%%%%%%%%%%%%%%%%%%%%%%%%%%

The frequency $x_{e,abs}$ is the maximum frequency at which %
the considered photon emitting processes could re-estabilish a thermal 
(blackbody) 
spectrum after the distortion. 
On the other hand, for 
$x_e \gsim x_{e,c}$ the timescale $t_C$ is shorter than
$t_{abs}$ and the high Compton scattering efficiency
prevents the formation of a blackbody spectrum and, as a particular case, 
for $z>z_1$
leads to a Bose-Einstein like spectrum with
a frequency dependent chemical potential 
(possibly evolving in time) tending to vanish at 
$x_e \ll x_{e,c}$ and to a value independent on the frequency
at $x_e \gg x_{e,c}$, because of the different relative relevance
of Compton scattering and photon emission processes.
Here $z_1$ is the redshift for which the dimensionless time variable 
$y_e(z)$ defined by
\begin{equation}
y_e(z) = \int^{t_0}_{t} {dt \over t_C}
=\int^{1+z}_{1} {d(1+z) \over 1+z} {t_{exp}\over t_C} \, 
\label{eq_y_e}
\end{equation}
assumes the value $y(z_1)\simeq\,5$ 
\cite{ZS69,illarionovsunyaev,chanjones,buriganaetal91a}
and the kinetic equilibrium between matter and radiation
can be reached.

In Fig.~\ref{fig:xc} we show the comparison between %
the dimensionless frequency $x_{e,CE}(z)$ and the other characteristic %
frequencies $x_{e,abs}(z)$ and $x_{e,c}(z)$ for different values 
of the energy %
injection rate $\Delta\varepsilon/\varepsilon_i$ and of the 
magnetic field $B_0$, %
in the cosmic epochs of interest here. As evident, 
$x_{e,CE} \ll x_{e,c}$
and $x_{e,CE} \ll x_{e,abs}$ at any time.
This implies that, 
in the presence of a dissipation mechanism operating to perturb a 
blackbody spectrum through a non vanishing Compton scattering term 
associated to a disequilibrium between matter and radiation 
temperatures, the cyclotron emission process term occurs at frequencies
where bremsstrahlung and double Compton are very efficient 
in keeping the radiation spectrum close to a blackbody 
at matter temperature.
In the next section we will quantitatively investigate this point. 

\section{Results}
\label{sec:results}

\subsection{Comparison between photon and energy injection rate}
\label{sec:confrontitrarates}

In this section we compare the contributions given by the different 
photon production terms in the 
Kompaneets equation for 
various cosmic epochs. 
We will compute the photon number production rate
\begin{equation}
\label{eq:ratesdingenerica}
\left(\frac{d n}{d t}\right)=
8\pi\left(\frac{k_BT_e}{h c}\right)^3\int 
\left(\frac{\partial\eta(x_e)}%
{\partial t}\right)x_e^2dx_e
\end{equation}
and the photon energy production rate 
\begin{equation}
\label{eq:ratesdiegenerica}
\left(\frac{d \varepsilon}{d t}\right)=
\frac{8\pi}{h^3 c^3}(k_BT_e)^4\int 
\left(\frac{\partial\eta(x_e)}%
{\partial t}\right)x_e^3dx_e \, .
\end{equation}
Using 
Eq. (\ref{eq:ratesdicy})
for the CE term we find
\begin{eqnarray}
\label{eq:productionrates_cy}
\left(\frac{d n}{d t}\right)_{CE}&=&8\pi
\left(\frac{k_BT_e}{hc}\right)^3\,K_{CE}(z)\int[1-\eta(x_e)(e^{x_e}-1)]
\delta(x_e-x_{e,CE})x_e^2dx_e \nonumber\\
\left(\frac{d \varepsilon}{d t}\right)_{CE}&=&\frac{8\pi}{h^3c^3}
(k_BT_e)^4\,K_{CE}(z)\int[1-\eta(x_e)(e^{x_e}-1)]
\delta(x_e-x_{e,CE})x_e^3dx_e \, .\nonumber\\
\mbox{ }
\end{eqnarray}   
Using the expressions given by Burigana et al. (1991) and
Burigana et al. (1995),
%\cite{buriganaetal91a,buriganaetal95} 
for the BR and DC term we have
\begin{eqnarray}
\label{eq:ratesdibredc}
\!\!\!\!\!\!\left(\frac{d n}{d t}\right)_{BR}\!\!&=&
K_{BR}(z)
\frac{8\pi k_B^3T_e^3}{c^3h^3}\int \frac{g_{BR}(x_e)}{x_e}
[1-\eta(x_e)(e^{x_e}-1)]\;e^{-x_e}dx_e \nonumber
\nonumber\\
\!\!\!\!\!\!\left(\frac{d \varepsilon}{d t}\right)_{BR}\!\!
&=&K_{BR}(z)
\frac{8\pi k_B^4T_e^4}{c^3h^3}\int\!g_{BR}(x_e)[1-\eta(x_e)(e^{x_e}-1)]\;
e^{-x_e}dx_e \, ,\nonumber\\
\\
\!\!\!\!\!\!\left(\frac{d n}{d t}\right)_{DC}\!\!&=&K_{DC}(z)
\frac{8\pi k_B^4T_e^3}{c^3h^3}\int \frac{[1-\eta(x_e)(e^{x_e}-1)]}{x_e}\;
 e^{-x_e/2}\;dx_e \nonumber\\
\nonumber\\
\!\!\!\!\!\!\!\!\left(\frac{d \varepsilon}{d t}\right)_{DC}\!\!&=&K_{DC}(z)
\frac{8\pi k_B^4T_e^4}{c^3h^3}\int [1-\eta(x_e)(e^{x_e}-1)]\;
e^{-x_e/2}\;dx_e \, ,\nonumber
\end{eqnarray}
where the double Compton Gaunt factor 
(in the limit of a BE-like spectrum; Burigana et al. (1995))
%\cite{buriganaetal95}) 
has been explicitely included.

\subsection{Early distortions}
\label{sec:earlydistortion}

We compare the contributions by the three radiative processes considered 
here for an energy dissipation occurring at high redshifts by assuming %
for $\eta$ (i.e. in
Eqs. (\ref{eq:productionrates_cy}) and (\ref{eq:ratesdibredc}))
a pure Bose-Einstein (BE) formula or a  %
BE-like spectrum with a frequency dependent chemical potential %
(Eqs. (\ref{eq:BE}) %
and (\ref{eq:chemicalpotential})), %
as expected under realistic conditions for the active phase of %
a dissipation process and the subsequent CMB spectrum evolution
\cite{sunyaevzeldovich,illarionovsunyaev}:
\begin{eqnarray}
\label{eq:BE}
\eta_{BE}=\frac{1}{e^{x_e+\mu(x_e)}-1} \, ,
\end{eqnarray}
\begin{eqnarray}
\label{eq:chemicalpotential}
\mu(x_e)=\mu_0 e^{-x_{e,c}/x_e} \, .
\end{eqnarray}
Neglecting the subsequent decreasing of the chemical potential 
due to the photon emission/absorption processes, 
the amount of fractional energy dissipated in the plasma 
for a process with a negligible photon production
is related to $\mu_0$ 
by a well known relation \cite{sunyaevzeldovich,DD77}
that in the limit of small distortions is simply expressed by 

\begin{eqnarray}
\label{eq:mudesue}
\mu_0 \simeq 1.4 \Delta \epsilon / \epsilon_i \, ,
\end{eqnarray}
where $\epsilon_i$ is the radiation energy density 
before the beginning of the process.

%%%%%%%%%%%%%%%%%%%%%%%Early distortions%%%%%%%%%%%%%%%%%%%%%%%%%%%%%%%%%%%%%%%%%
%Figura 1
\linespread{1}
\begin{figure}[!th]
%\special{psfile=WMAP_ratesbe_d2.ps  hscale=75 vscale=75  hoffset=-0 voffset=-400}
\includegraphics{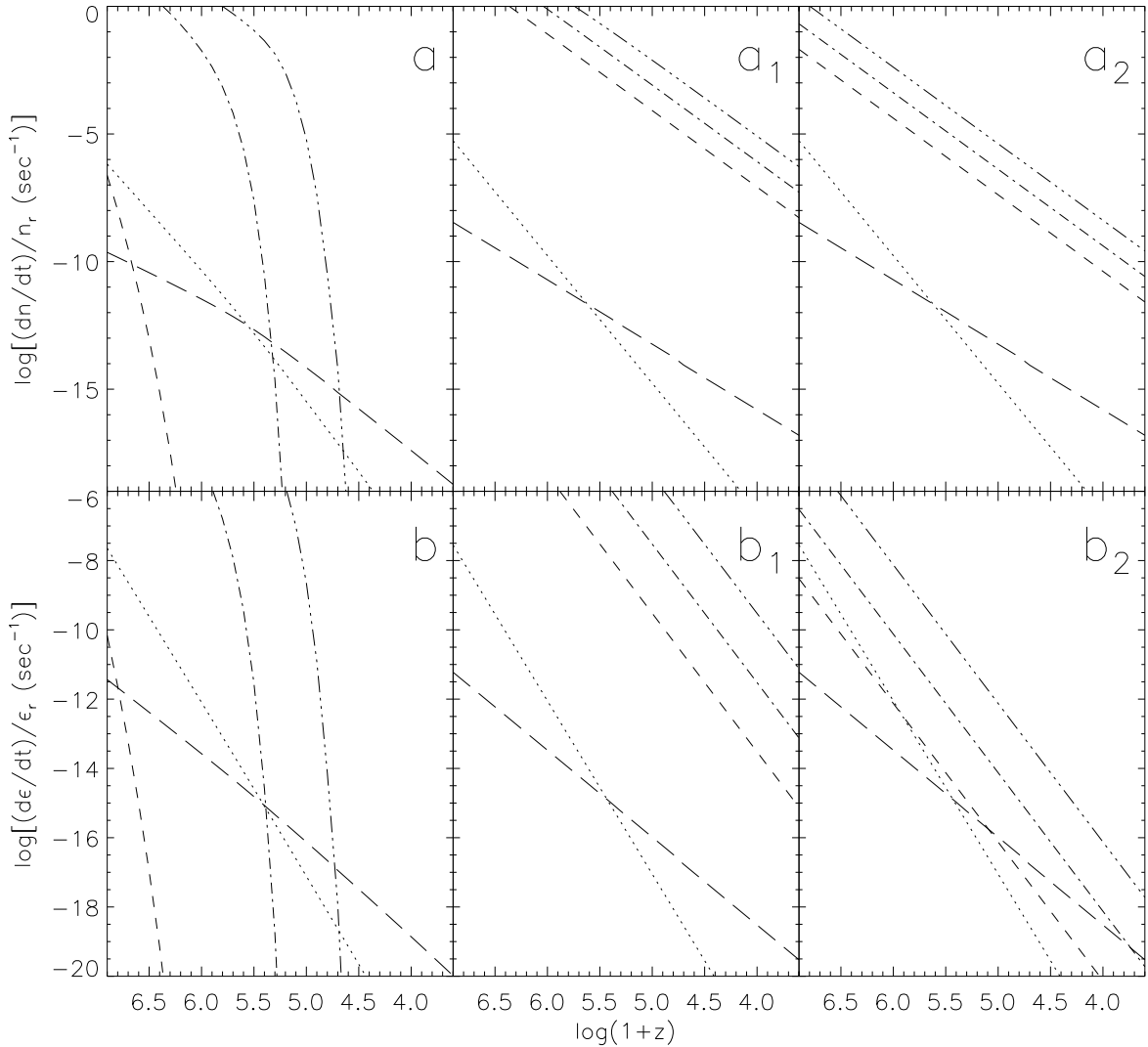}
\vspace{14. cm} 
\caption{%
Photon number (panels $a$ and $a_1$) 
and associated energy injection (panels $b$ and $b_1$) 
fractional (i.e. divided respectively by the photon number and energy 
density of a blackbody at temperature $T_0(1+z)$) 
rates at the redshift $z$
from DC (dots), BR (long dashes)
and CE 
(dashes: $B_0=2\times10^{-6}$~G;
dots-dashes: $B_0=2\times10^{-5}$~G;
three dots-dashes: $B_0=2\times10^{-4}$~G) in the presence of
an early distortion with $\mu_0= 1.4 \times 10^{-2}$
occuring exactly at the redshift $z$ (properly speaking this result
holds for $z \gsim z_1$ while it is only indicative for 
BE like distorted spectra at $z < z_1$ resulting from processing
occurred at $z \gsim z_1$). 
The panels $a$ and $b$ refer to computations 
including the proper frequency dependence of the chemical potential
[i.e. a BE-like spectrum, $\mu = \mu (x_e)$], 
neglected for comparison in panels 
$a_1$ and $b_1$ (i.e. $\mu = \mu_0$).
Panels $a_2$ and $b_2$ are identical to panels 
$a_1$ and $b_1$ but for 
$B_0=10^{-9}$~G (dashes),
$B_0=10^{-8}$~G (dots-dashes),
$B_0=10^{-7}$~G (three dots-dashes) and show that neglecting
the frequency dependence of the chemical potential
would imply the (wrong) conclusion of a dominance 
of the CE contribution for reasonable values of the magnetic
field. Note that the DC and BR rates in panels $a_1$ and $a_2$ 
are only indicative (see footnote 3).
The values used for the cosmological parameters 
are the same as in Fig.~\ref{fig:xc}.
%[We assume here $\Omega_b=0.044$, $H_0=71$, $z_{rec}=10^3$, 
%$\Omega_{tot}=1$].
}
\label{fig:rates_be_DE/E_I=d-4}  
\end{figure}
%Figura 3
%\begin{figure}[h]
\begin{figure}[!th]
%\special{psfile=WMAP_ratesbe_d0.ps  hscale=75 vscale=75  hoffset=-0 voffset=-400}
\includegraphics{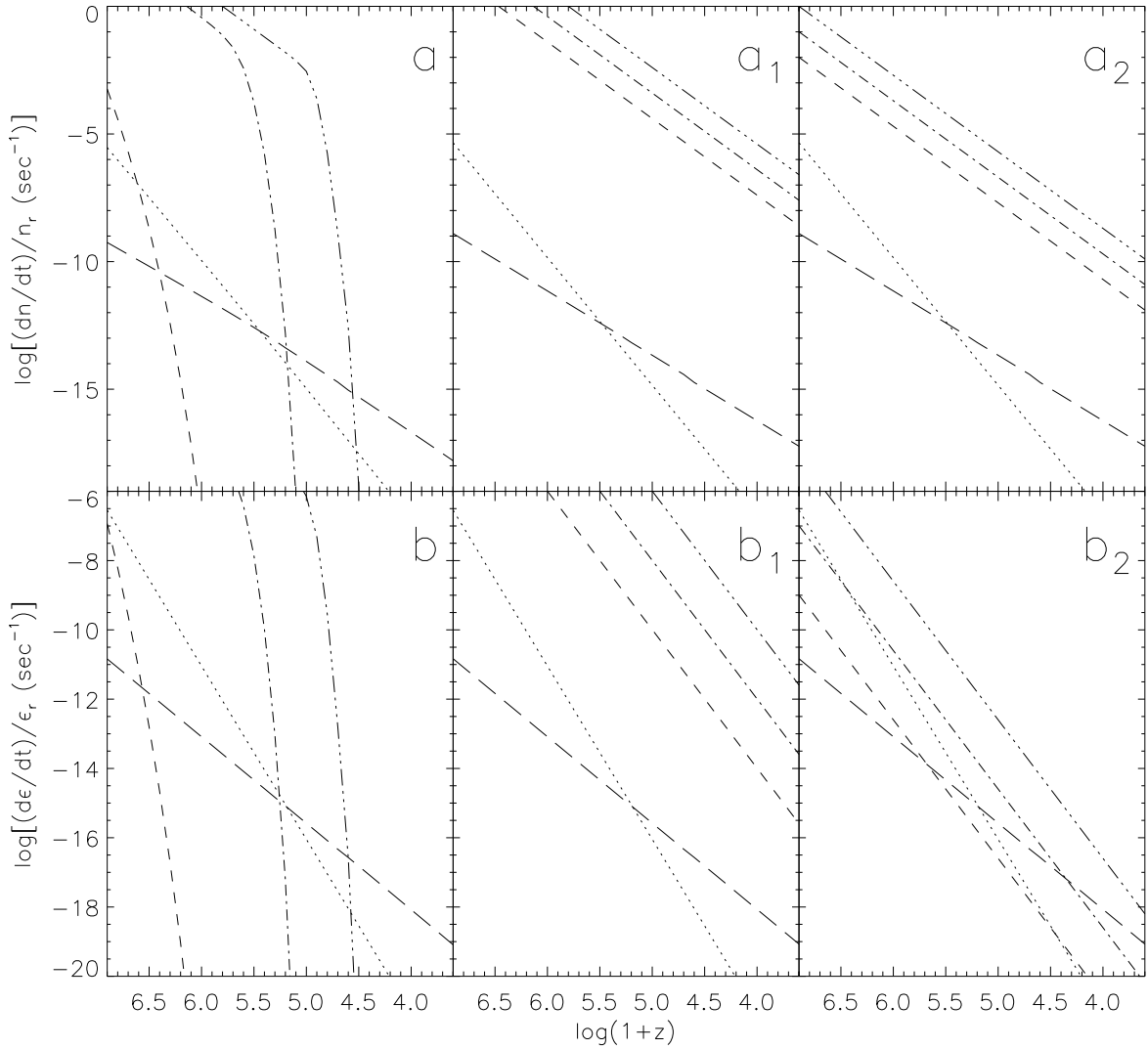}
\vspace{14. cm} 
\caption{The same as in Fig.~\ref{fig:rates_be_DE/E_I=d-4} but for 
$\mu_0=1.4$.}
\label{fig:rates_be_DE/E_I=d0}  
\end{figure}
%%%%%%%%%%%%%%%%%%%%%%%%%%%%%%%%%%%%%%%%%%%%%%%%%%%%%%%%%%%%%%%%%%%%%%%%%%%%%%%
The result of our computation is reported in 
Figs.~\ref{fig:rates_be_DE/E_I=d-4} 
and \ref{fig:rates_be_DE/E_I=d0} 
for a representative choice of
the cosmological parameters and some values of the cosmic 
magnetic field~\footnote{Note that while all the integrals  
in Eq.~(\ref{eq:ratesdibredc}) converge and 
do not depend on the low frequency 
cut-off, $x_l$, adopted in the numerical integration
(provided that $x_l \ll x_{e,c}$), 
in the case of a photon distribution function described by 
the pure BE formula
the photon number production rates do not converge
because of the low frequency $1/x_e$ dependence. 
The results reported in panels $a_1$ and $a_2$ 
of Figs.~\ref{fig:rates_be_DE/E_I=d-4} 
and \ref{fig:rates_be_DE/E_I=d0} 
for the DC and BR 
photon number production rates, 
obtained by using ${\rm log} x_l \simeq -4.3$,
are only indicative.
Of course, this does not affect the conclusion 
of our work that is based on the proper BE-like spectrum
which frequency dependent chemical potential
removes the above divergency (see panel $a$ 
of Figs.~\ref{fig:rates_be_DE/E_I=d-4} 
and \ref{fig:rates_be_DE/E_I=d0}).}.

Considering the pure BE approximation, exploited for example by 
Afshordi (2002), and neglecting the consideration in footnote 3,
%\cite{afshordi}) 
the CE contributions appear to be much larger than the contributions by 
BR and DC even for 
$B_0\lsim 10^{-8}$, i.e. for realistic values of magnetic fields
(see panels $a_1,\,a_2,\,b_1\mbox{ and }b_2$ 
in Figs.~\ref{fig:rates_be_DE/E_I=d-4} and 
\ref{fig:rates_be_DE/E_I=d0}). 
Since this result is obtained under the assumption that the chemical 
potential $\mu$ is constant with respect to the frequency it 
neglects at all the efficiency of bremsstrahlung and double Compton
at low frequencies in the formation of the distorted spectrum
during the energy dissipation phase. 

In reality, the result is very different when we adopt
a better approximation of the distorted spectrum, represented
by the BE-like spectrum as discussed above.
The differences found for CE term
by assuming a constant or a frequency dependent chemical potential
are much larger than those %
found for the DC and BR terms 
(see Figs.~\ref{fig:rates_be_DE/E_I=d-4} and \ref{fig:rates_be_DE/E_I=d0})
because of the different %
frequency location of their contributions (see Fig.~\ref{fig:xc}).
Replacing 
the expression 
for $\eta$ 
given in Eqs. (\ref{eq:BE}) %
and (\ref{eq:chemicalpotential}) 
in Eqs. (\ref{eq:productionrates_cy}) %
one obtains in fact
\begin{eqnarray*}
\left(\frac{d n}{d t}\right)_{CE}\!\!\!&=&8\pi
\left(\frac{k_BT_e}{hc}\right)^3\,K_{CE}(z)\left\{1-\frac{\exp (x_e)-1}{\exp[x_e+\mu_0
e^{(-x_{e,c}/x_{e,CE})}]-1}\right\}x_{e,CE}^2\\
\left(\frac{d \varepsilon}{d t}\right)_{CE}\!\!\!&=&\frac{8\pi}{h^3c^3}
(k_BT_e)^4\,K_{CE}(z)\left\{1-\frac{\exp (x_e)-1}
{\exp[x_e+\mu_0e^{(-x_{e,c}/x_{e,CE})}]-1}\right\}x_{e,CE}^3 \, .\\
\end{eqnarray*}

\begin{figure}[!th]
%\special{psfile=vedi_B0_VS_Energy.ps  hscale=62 vscale=62  hoffset=10 voffset=-360}
\includegraphics{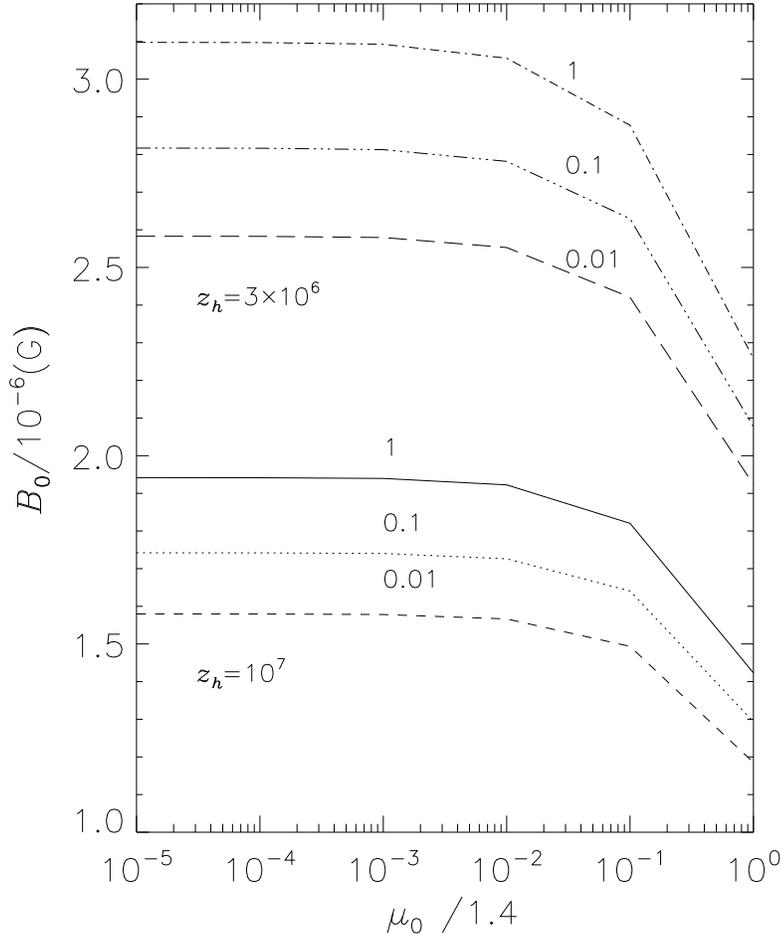}
\vspace{12.5 cm} 
\linespread{1}
\caption{Magnetic field $B_0$ 
for which the ratio between the photon number production rates by 
cyclotron 
emission and bremsstrahlung plus double Compton
assumes the values indicated close to each curve (1, 0.1, 0.01)
as a function of the 
of fractional injected energy.
The values used for the cosmological parameters 
are the same as in Fig.~\ref{fig:xc}
(see also the text).} 
\label{fig:B_0_VS_ENERGY}  
\end{figure}

Since for $z>z_1$ it is $x_{e,CE}\ll x_{e,c}$ (see Fig.~\ref{fig:xc}), %
from Eq. (\ref{eq:chemicalpotential}) %
it is straightforward to obtain 
$\mu(x_{e,CE})\ll 1$ (for $B_0\lsim 10^{-6}$), and the above rates  
become then very small or negligible \cite{zizzo03}.
We find in fact that the CE term is comparable %
to the BR and DC term only for values of $B_0$ larger %
than $\sim 10^{-6}$~G (see Fig.~\ref{fig:B_0_VS_ENERGY}
and the following discussion), the exact value 
depending on 
$\Delta\varepsilon/\varepsilon_i$ and on the cosmological parameters.
This value of $B_0$ is much larger than the values obtained from  
observations in IGM %
(see Sect.~\ref{sect:osservazioni_di_B}) and of the predictions of the 
scenarios %
for the genesis of cosmic magnetic fields (see 
Sect.~\ref{sect:genesisofB}). 

Differently from previous analyses, we conclude that the CE contribution 
to the thermalization process of CMB spectrum
after an early heating is negligible.

For sake of generality, we have computed the value of the cosmic magnetic 
field for which the CE photon production rate is equal to 
(or is 1/10 or 1/100 of)
the combined photon production rate by BR and DC. 
We report in Fig.~\ref{fig:B_0_VS_ENERGY} 
the results found at the two 
representative
redshifts, $z=10^7$ and $z=3\times 10^6$, quite close to the 
thermalization redshift, 
since the relative contribution by the CE term 
to the global photon production 
is more relevant at higher redshifts 
(see Figs.~\ref{fig:rates_be_DE/E_I=d-4} %
and \ref{fig:rates_be_DE/E_I=d0}). 
As evident from Fig.~\ref{fig:B_0_VS_ENERGY}, 
for $\Delta \epsilon / \epsilon_i \lsim {\rm some} \times 10^{-2}$
the result is very weakly dependent on $\Delta \epsilon / \epsilon_i$
while it significantly depends on the assumed ratio
between the CE rate and the BR and DC combined rate.
Exploiting the results at 
$\Delta \epsilon / \epsilon_i \lsim {\rm few} \times 10^{-2}$,
we have then searched for a fit of $B_0$ as a function of the ratio of 
these photon production rates 
in the form: 
\begin{equation}\label{ratioph}\nonumber
{B_0 \over 10^{-6}{\rm G}}=\alpha\log\left(\frac
{\dot{n}_{CE}}{\dot{n}_{BR}+\dot{n}_{DC}}\right) + \beta\, ,
\end{equation}
where the overdots denote the time derivatives.
The best-fit values of the 
parameters are
$\alpha \simeq 0.178$ and 
$\beta \simeq 1.924$ for  $z_h=10^7$
and 
$\alpha \simeq 0.252$ 
and $\beta \simeq 3.064$
for $z_h=3\times10^6$.

%%%%%%%%%%%%%%%%%%%%%%%%%%%%B_0_VS_ENERGY%%%%%%%%%%%%%%%%%%%%%%%%%%%%%%%%%%%%%%
%%%%%%%%%%%%%%%%%%%%%%%%%%%%%%%%%%%%%%%%%%%%%%%%%%%%%%%%%%%%%%%%%%%%%%%%%%%%%%%
%
%%%%%%%%%%%%%%%%%%%%%%%%%%%%B_0_VS_ENERGY%%%%%%%%%%%%%%%%%%%%%%%%%%%%%%%%%%%%%%
%\begin{figure}[!h]
%\special{psfile=WMAP_bvsrates.ps hscale=62 vscale=62  hoffset=10 voffset=-360}
%\vspace{12.5 cm} 
%\linespread{1}
%\caption{We plot here the best fit of magnetic field $B_0$ corresponding 
%to several values of $log\left(\frac{(dn/dt)_{CE}}{(dn/dt)_{dc}+(dn/dt)_{br}}\right)$. %
%There have been considered two values of $z_h$,
%the redshift at wich the heating occurs %
%($3\times 10^6$ thick dashed line, $10^7$ thick solid line).} 
%\label{fig:B_0_VS_RATES}  
%\end{figure}
%%%%%%%%%%%%%%%%%%%%%%%%%%%%%%%%%%%%%%%%%%%%%%%%%%%%%%%%%%%%%%%%%%%%%%%%%%%%%%%
%

\subsection{Late distortions}
\label{sec:latedistortions}

In the case of small, late distortions ($y_h \ll 1$), 
a quite good approximation %
of the high frequency region of the CMB spectrum
(observationally, at $\lambda\lsim 1$~cm), where
Compton scattering dominates, is provided by the 
well known solution \cite{ZS69}
\begin{eqnarray}\label{eq:solbassiz_H}
\eta(x,\tau) \simeq  
\eta^H(x,\tau) 
= \eta_i+u\frac{x/\phi_i\exp(x/\phi_i)}
{[\exp(x/\phi_i)-1]^2}\left(\frac{x/\phi_i}{\tanh(x/2\phi_i)}-4\right) \, 
,
\end{eqnarray}

where $\eta_i$ is the initial distribution function.

Here $u$ is the (evolving) Comptonization parameter
\begin{eqnarray}\label{eq:upar1}
u(t)=u(z) = \int_{t_i}^t\frac{\phi-\phi_i}{\phi}\frac{dt}{t_C}=
\int_{1+z}^{1+z_i}(\phi-\phi_i)\left(\frac{k_BT_r}{m_ec^2}\right)n_e\sigma_T
t_{exp}\frac{d(1+z^{\prime})}{1+z^{\prime}} \, .
\end{eqnarray}

By integrating $u$ over the whole relevant energy dissipation phase
we have the ``usual'' Comptonization parameter
related to the whole fractional energy exchange by the 
well known expression \cite{zeldovich} 
\begin{equation}
u\simeq (1/4)\Delta\varepsilon/\varepsilon_i \, .
\end{equation}

In the case of an initial blackbody spectrum $\eta_i$ is given by
\begin{equation}\label{eq:eta_i}
\eta_i=\eta_{BB}=\frac{1}{e^{x/\phi_i}-1} \, ,
\end{equation}

where

\begin{equation}\label{eq:fi_i}
\phi_i=(1+\Delta\varepsilon/\varepsilon_i)^{-1/4}\simeq 1-u \, 
\end{equation}

is the ratio between electron and radiation temperature
before the beginning of the process
(only small modifications of above formulas 
are needed for initial BE like spectra; see 
Burigana et al. (1995)).
%\cite{buriganaetal95}).

For $z < {\rm min}(z_p, \tilde{z})$, with $\tilde{z} = {\rm min} 
(z_1,z_h)$,
$z_p$ being the redshift~\footnote{$\kappa$ 
($\simeq 1.68$ for 3 species of massless neutrinos)
takes into account the
contribution of relativistic neutrinos to the dynamics of the
universe. Strictly speaking the present ratio of neutrino to
photon energy densities, and hence the value of $\kappa$, is itself a
function of the amount of energy dissipated. The effect, however,
is never very important and is negligible
for very small distortions.}
$z_p=2.14 \times 10^4 (T_0/2.7K)^{1/2}\Ohat^{-1/2}
(\kappa/1.68)^{1/4}$
\cite{buriganaetal91a,buriganaetal95},
%the dimensionless frequency $x_{e,abs}(z)$ decrease with $z$, %
%then $x_{e,abs}(z)$ is the maximum frequency, for $z<z_p$, at which a %
%blackbody spectrum can be re-estabilished after the generation of a 
%spectral distortion.
the CMB spectrum at very low frequencies 
(observationally, at $\lambda\gsim {\rm some}$~dm)
is mainly determined by the BR term (the DC efficiency significantly 
decreases at decreasing redshifts)
and it can be a quite well approximated by the expression
\cite{danesedezotti,buriganaetal91a,buriganaetal95}

\begin{eqnarray}\label{eq:solbassiz_L}
\eta(x,\tau) & \simeq & \eta^L(x,\tau) \nonumber \\ 
& = & \eta_i\exp[-(\tau-\tau_h)]+\exp[-(\tau-\tau_h)]
\int_{\tau_h}^{\tau}{\exp(\tau^{\prime}-\tau_h)
\over \exp[x/\phi(\tau^{\prime})]-1}d\tau^{\prime}
\, ,
\end{eqnarray}

where a constant value of $\phi(\tau)$, 
related to the value of the fraction of injected energy, 
$\Delta\varepsilon/\varepsilon_i$,
has been assumed for simplicity, in order
to make the treatment fully analytical.

At intermediate frequencies both BR and C are important and
the spectrum is well described by the expression \cite{buriganaetal95}
%\begin{eqnarray}\label{eq:solbassiz}
$$\eta(x,\tau) \simeq \eta^L(x,\tau) + \eta^H(x,\tau) -\eta_i \, .$$
%\end{eqnarray}

\begin{figure}[!th]
\vskip 1cm
%\special{psfile=WMAP_Vedi_late_distortion.ps  hscale=75 vscale=75  hoffset=-15 voffset=-400}
\includegraphics{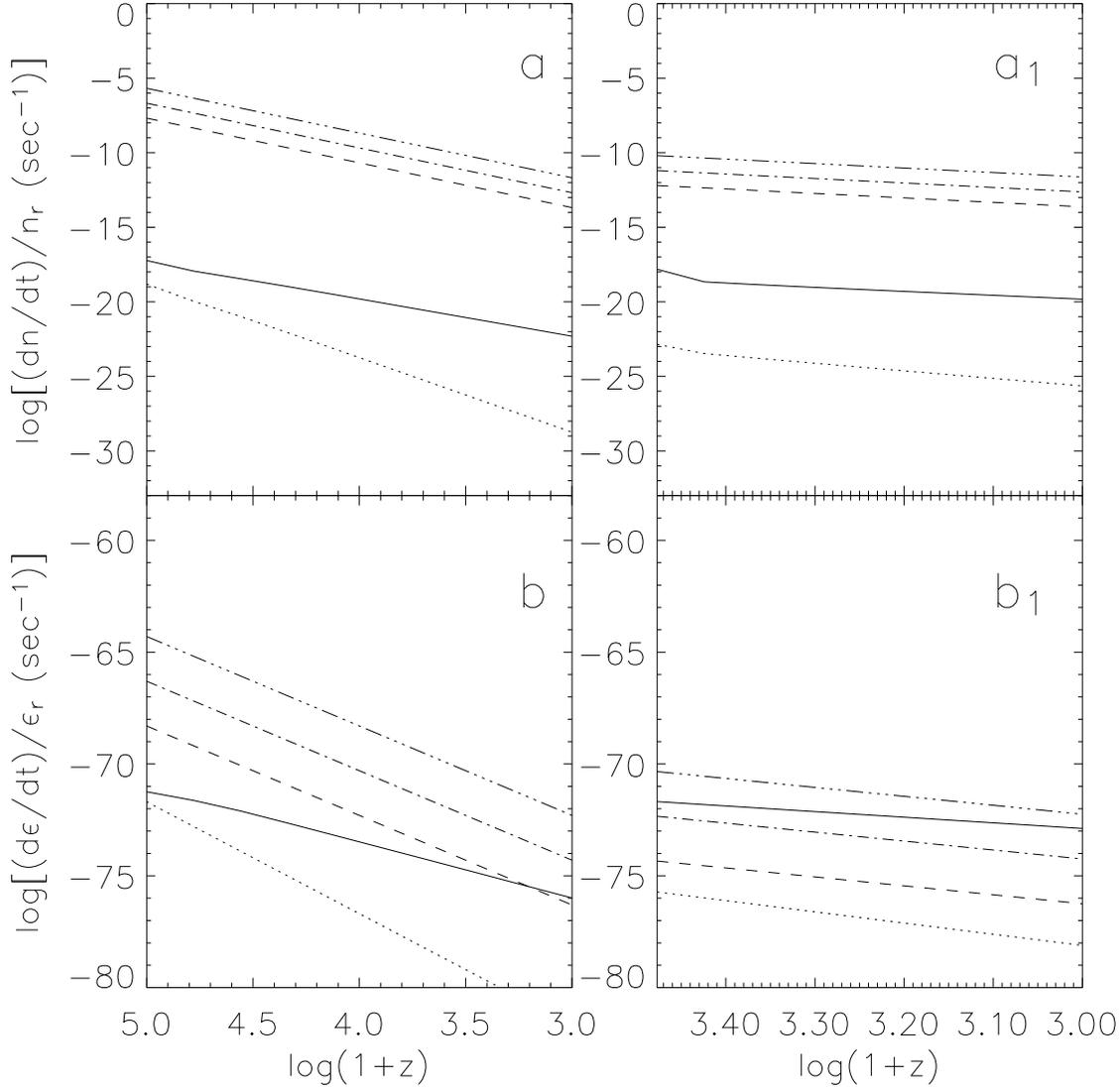}
\vspace{14.3 cm} 
\linespread{1}
\caption{Photon production rates (panels $a$ and $a_1$) 
and energy production rates %
(panels $b$ and $b_1$) for the three processes taken into account. 
The expression adopted %
for $\eta$ is given by the Eq.~(\ref{eq:solbassiz_H}). 
We have assumed here:
$\Delta\varepsilon/\varepsilon_i=10^{-4}$; 
$B_0=10^{-9}$ (short dashed 
line),
$B_0=10^{-8}$ (dot dashed line), and $B_0=10^{-7}$~G (three dot dashed 
line);
$z_h=10^5$ (panels $a$ and $b$) 
and $z_h=3\times 10^3$ (panels $a_1$ and $b_1$).
Thick solid line: double Compton scattering rates; 
dots: bremsstrahlung rates. 
The values used for the cosmological parameters 
are the same as in Fig.~\ref{fig:xc}.
}
\label{fig:CO_d-4_d5_987}  
\end{figure}

In Fig.~\ref{fig:CO_d-4_d5_987} we compare the rates of CE with those %
of BR and DC using for $\eta$ the expression in 
Eq.~(\ref{eq:solbassiz_H}), %
which, as already mentioned, is a good approximation for the
high frequency range and describes the spectrum when only Compton
scattering 
operates~\footnote{Note the analogy with the result obtained in the 
limit of a pure BE spectrum in the case of early dissipation processes.}. 

%%%%%%%%%%%%%%%%%%%%%Figura Comptonizzazione%%%%%%%%%%%%%%%%%%%%%%%%%%%%%%%%%%%
%%%%%%%%%%%%%%%%%%%%%%%%%%%%%%%%%%%%%%%%%%%%%%%%%%%%%%%%%%%%%%%%%%%%%%%%%%%%%%%

However, in the frequency interval %
where CE is located (i.e. $x_e\ll 1$) the spectrum is %
well approximated by Eq. (\ref{eq:solbassiz_L}). %
Assuming a constant value of $\phi(\tau)$ 
implies that we can easily perform the integral in 
Eq.~(\ref{eq:solbassiz_L}) to have%
\begin{eqnarray}
\label{eq:bassizapprossimata}
\eta(x,\tau)&=&\eta_i\exp[-(\tau-\tau_h)]+\frac{1}{e^{x/\phi_f}-1}-
\frac{\exp[-(\tau-\tau_h)]}{e^{x/\phi_i}-1} \nonumber \\
&=&\eta_i e^{-y_{abs}}+\frac{1}{e^{x/\phi_f}-1}-
\frac{e^{-y_{abs}}}{e^{x/\phi_i}-1} \, ,
\end{eqnarray}
where $\phi_f=\phi(\tau)$
and Eq.~(\ref{eq:defxabs}) has been used in the last equality.
%
%
%
%%%%%%%%%%%%%%%%%%%%%%%FIG.-Y_abs(x_e,cy)%%%%%%%%%%%%%%%%%%%%%%%%%%%%%%%%%%%%%%
%\begin{figure}[!htbp]
\begin{figure}[!th]
%\special{psfile=WMAP_ystep.ps  hscale=62 vscale=62  hoffset=20 voffset=-395}
\includegraphics{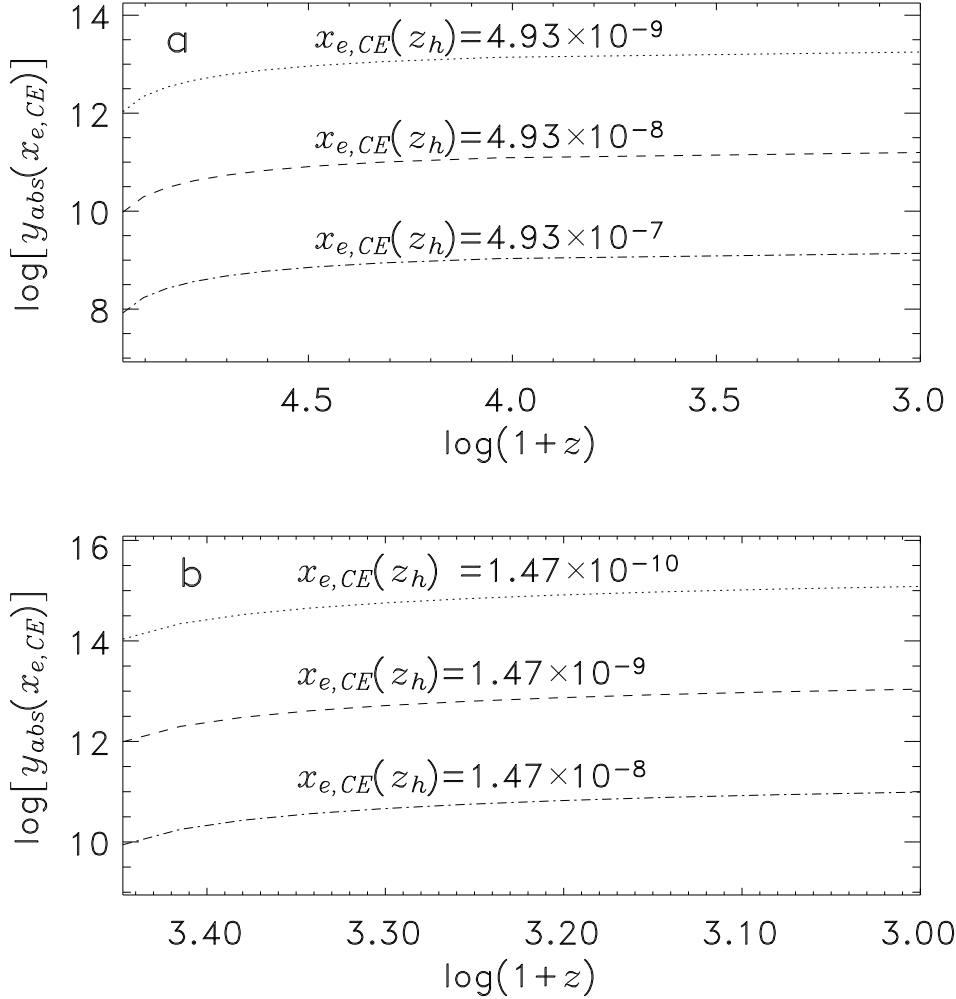}
\vspace{13.3 cm} 
\linespread{1}
\vskip 0.5cm
\caption{Optical depth $y_{abs}$ at the frequency 
$x_{e,CE}(z_h)$ (reported close to each curve)
evaluated by computing the integral in 
Eq.~\ref{eq:defxabs} but within $1+z$ and $1+z_h$,
as function of $z$.
We assume $z_h=10^5$ in panel $a$ %
and $z_h=3\times 10^3$ in panel $b$.
Again $B_0=10^{-9}$, 
$B_0=10^{-8}$,
and $B_0=10^{-7}$~G (from the top to the bottom).
Note that keeping $x_{e,CE}$ at the value it has at 
the largest redshift in the interval $[z,z_h]$
implies an overestimation of $x_{e,CE}$ and then
an underestimation of the photon production 
by bremsstrahlung and double Compton.
The values of $y_{abs}$ 
reported in the figure have then to be 
considered as lower limits.
The values used for the cosmological parameters 
are the same as in Fig.~\ref{fig:xc}.
}
\label{fig:yabsstep}  
\end{figure}
%\begin{figure}[!htbp]
\begin{figure}[!th]
%\special{psfile=WMAP_yabsinf.ps  hscale=62 vscale=62  hoffset=20 voffset=-395}
\includegraphics{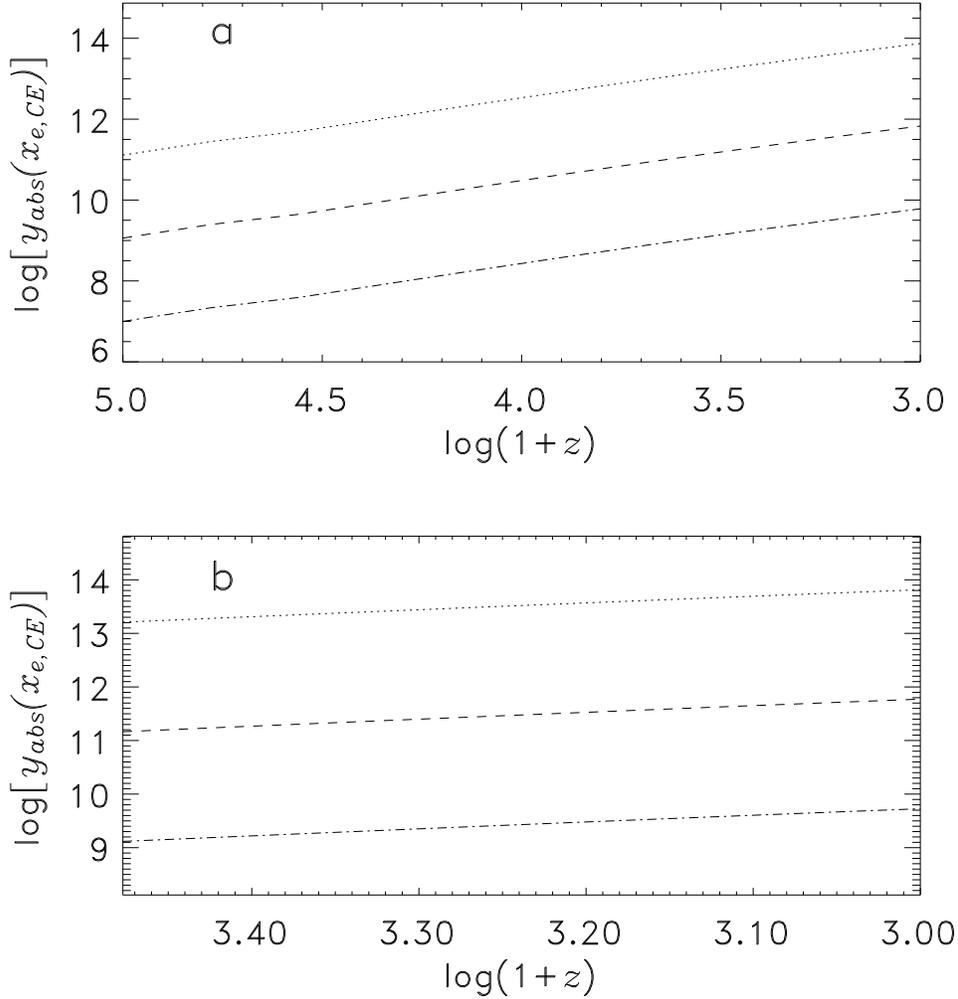}
\vspace{13.3 cm} 
\linespread{1}
\vskip 0.5cm
\caption{Optical depth $y_{abs}$ at the frequency $x_{e,CE}(z)$
evaluated by computing the integral in 
Eq.~(\ref{eq:defxabs}) but within a small interval $\Delta z$ 
($\simeq z/100$) about $z$,
as function of $z$, for the same values of $B_0$ as in 
Fig.~\ref{fig:yabsstep}.
Analogously to Fig.~\ref{fig:yabsstep}, we 
report in two distinct panels the case at $z$ between
$10^5$ and $10^3$ 
(panel $a$) 
and the case at $z$ between
$3\times 10^3$ and $10^3$
(panel $b$; in this context, it is in practice a zoom up
of panel $a$).
The values used for the cosmological parameters 
are the same as in Fig.~\ref{fig:xc}.
}
\label{fig:yabsinf}  
\end{figure}

%%%%%%%%%%%%%%%%%%%%%%%%%%%%%%%%%%%%%%%%%%%%%%%%%%%%%%%%%%%%%%%%%%%%%%%%%%%%%%%
From Fig.~\ref{fig:xc} it is clear that for $z~<~z_1$ we have %
$x_{e,CE}(z)\ll x_{e,abs}(z)$, then $y_{abs}(x_{e,CE})\gg 1$ %
(see Figs.~\ref{fig:yabsstep} and \ref{fig:yabsinf})
and consequently $e^{-y_{abs}}\rightarrow 0$;  
Eq.~(\ref{eq:bassizapprossimata}) 
reduces then to 
\begin{eqnarray*}
\eta(x_e,\tau)\rightarrow\frac{1}{e^{x/\phi_f}-1} \, .
\end{eqnarray*}
Replacing this expression for $\eta$ in Eq.~(\ref{eq:ratesdicy}) gives %
\begin{eqnarray*}
\left(\frac{\partial\eta}{\partial t}\right)_{CE} 
\simeq \frac{4\pi^2e\,c\,n_e}%
{3B(z)}\,[1-\frac{1}{e^{x/\phi_f}-1}\,(e^{x/\phi_f}-1)]\,%
\delta(x/\phi_f-x_{CE}/\phi_f)\rightarrow0 \, .
\end{eqnarray*}
This result, that can be generalized to thermal histories 
with variable $\phi(\tau)$, 
is a consequence of the fact that for 
$z<z_p$ photons of frequency $x_e<x_{e,abs}$ %
are absorbed before Compton scattering can efficiently shift 
them to higher frequencies during the active phase of the
dissipation process, as
illustrated by Fig.~\ref{fig:yabsinf}:
even integrated over a small interval
in redshift, taken sufficiently 
small to assure that $\phi(\tau)$ could be considered as a constant
in practice within the interval (as possible for reasonably smooth thermal 
histories with variable $\phi(\tau)$),
the (bremsstrahlung dominated at these redshifts) 
optical depth $y_{abs}$ 
at the (very low) cyclotron frequency
is so large to keep the spectrum very close to a blackbody 
at electron temperature at each time.

\section{Discussion and conclusion}
\label{conclusion}

In this paper 
we have investigated 
the role of the cyclotron emission associated to cosmic magnetic fields
on the evolution of CMB spectral distortions. 
We properly included the contributions by spontaneous and stimulated 
emission and by absorption. 
We have derived 
the photon and energy injection rates due to cyclotron emission
exploiting and generalizing the 
approach by Afshordi (2002).
%\cite{afshordi}.
%approach by Afshordi \cite{afshordi}.
The photon and energy production rates by cyclotron emission
have been numerically compared 
with those of the relevant radiative processes
operating in the cosmic plasma, 
bremsstrahlung and double Compton scattering.
We have evaluated the
role of cyclotron emission 
by adopting realistic approximations for the CMB distorted spectra.
In the case of early distorted spectra
we adopted a Bose-Einstein distribution function 
with a frequency dependent chemical potential, 
as realistically expected by an early 
dissipation process.
For realistic values of $B_0$, consistent with observations, 
we find that the
cyclotron emission term is negligible with respect to that 
from bremsstrahlung and double Compton, because of 
the different frequency location of their contributions
and the very high efficiency of bremsstrahlung  and double Compton to keep 
the CMB spectrum 
very close to a blackbody 
shape (at electron temperature) during the  
spectral distortion formation.
 
We have also considered the role of cyclotron emission for a 
dissipation process possibly occurred at low redshift ($z_h\lsim z_1$)
by exploiting the available analytical description of late
distorted spectra.
Again the very low frequency location of the cyclotron emission
implies that it could be relevant only where bremsstrahlung and 
double Compton
very efficiently keep the blackbody shape of the CMB spectrum
during the active phase of the dissipation process
and the subsequent evolution.
Again, the cyclotron contribution is found to be negligible. 

Of course, our approach does not apply to possible
specific mechanisms of generation of spectral distortions 
in which the energy is injected in the cosmic radiation field
through a relevant modification of the photon occupation
number at very low frequencies, close to the cyclotron
emission frequency. A specific (typically, highly model dependent) 
treatment is needed in this case.

Differently from previous analyses, we then conclude that for a very 
large set of dissipation mechanisms, as in the case of 
dissipation processes mediated 
by energy exchanges between matter and radiation 
or associated to photon injections at frequencies significantly
different from the cyclotron emission frequency, 
the role of cyclotron emission 
in the evolution of CMB spectral distortions
is negligible for any realistic value of the cosmic magnetic field.
In particular, it cannot re-establish a blackbody
spectrum after the generation of a realistic early distortion
and it cannot produce a significant spectral distortion
(and therefore no significant limits on the cosmic magnetic
field strength can be set by constraints on CMB spectral distortions 
when interpreted as produced by cyclotron emission).

Consequently, 
the constraints on the energy dissipations at various
cosmic times set by the currently available data 
\cite{SB02}
and 
expected by future experiments \cite{KOG96,FM02,BS03a,Brev04}
can be derived, under quite general assumptions,
by considering only Compton
scattering, bremsstrahlung, and double Compton, other than,
obviously, the considered dissipation process(es).

As a particular application of CMB spectrum analyses
to cosmic magnetic fields, 
Jedamzik et al. (2000)
%\cite{jedamziketal2000} 
estimated 
a (weak) limit on the magnetic field 
strength $B_0 \lsim 10^{-8}$~G
on a coherence length of $\sim 400$~pc 
by comparing the constraints on CMB Bose-Einstein like distortions
(attenuated by taking into account the dominant double
Compton process~\footnote{Note that $10^{-8}$~G is a value much smaller
than that for which the role of cyclotron emission
is comparable to that of double Compton scattering (see 
Fig.~\ref{fig:B_0_VS_ENERGY}).}) 
with the distortion level predicted by the damping of 
magnetic field normal oscillation modes before recombination,
which associated dissipation rate is proportional to
the square of the magnetic field strength.

An interesting aspect
is the evaluation of the polarization degree
imprinted by the cyclotron emission on the CMB. While the (very small) 
effect on CMB spectral distortions shows a strong dependence only on the 
magnetic field strength, the polarization imprint depends in principle
also on the uniformity and structure of the magnetic field and, from
the observational point of view, on the experiment angular resolution
because of the obvious decreasing of the global polarization 
signal when it is measured within a solid angle larger than the 
field coherence angular scale. A detailed study, out of the 
scope of the present work, is therefore model dependent.
On the other hand, an approximate upper
limit to the polarization degree imprinted by the cyclotron emission on 
the CMB can be derived by exploiting Eq.~(\ref{ratioph}) at the
representative redshift $z \sim 3 \times 10^6$, i.e at an 
epoch significantly later
than the thermalization redshift of an arbitrarily large 
spectral distortion but before the strong decreasing of the
cyclotron photon production efficiency (see Fig.~\ref{fig:xc}
and panels $a)$ of Figs.~\ref{fig:rates_be_DE/E_I=d-4}
and \ref{fig:rates_be_DE/E_I=d0}). 
Assuming that after an energy  dissipation
producing a distorted spectrum with a chemical 
potential~\footnote{Only small distortions are compatible with current 
data at $z \sim 3 \times 10^6$  (Salvaterra and Burigana, 2002).}
$\mu_0 \simeq 1.4 \Delta \epsilon/\epsilon_i$ the photon
production processes are able to fully re-establish a Planckian spectrum,
the additional photon fractional number is 
$\Delta n/n_r \simeq [ n_P(T_r) - n_P(T_r) \phi_{BE}^3(\mu_0) \varphi(\mu_0) ] 
 / n_P(T_r)$; since  
$\phi_{BE}(\mu_0) \simeq (1-1.11\mu_0)^{-1/4}$ 
and $\varphi(\mu_0) \simeq 1-1.368\mu_0$ we obtain
$\Delta n/n_r \simeq 0.779 \Delta \epsilon/\epsilon_i$.
From Eq.~(\ref{ratioph}) 
%using values of $\alpha$ and $\beta$
%evaluated at $z \sim 3 \times 10^6$,
we have that only a fraction 
$r \simeq 6.94 \times 10^{-13} \times 10^{3.97 (B_0/10^{-6}{\rm G})}$
of the above additional photons is produced by cyclotron emission.
Assuming an intrinsic polarization degree $f_p^i$ 
for the cyclotron emission from each coherence cell, 
a beam depolarization factor $f_p^{bd}$ due to 
the averaging of polarization signals with different 
orientations within the experiment beam size, 
a Faraday depolarization factor $f_p^{Fd}$, and
a polarization decreasing factor $f_p^{sc}$
due to the randomization of the polarization orientations
when (because of Compton scattering)
cyclotron emission photons migrate from frequencies 
$\approx \nu_c$ to higher frequencies where CMB experiments
are carried out, we derive
an upper limit of 
$d \sim 5.41 \times 10^{-13} (f_p^i f_p^{bd} f_p^{Fd} f_p^{sc}) 
(\Delta \epsilon/\epsilon_i)
\times 10^{3.97 (B_0/10^{-6}{\rm G})}$
for the contribution of cyclotron emission to the CMB polarization 
degree.
Even assuming $f_p^i f_p^{bd} f_p^{Fd} f_p^{sc} \sim 1$,
$\Delta \epsilon/\epsilon_i \sim 0.1$,
and $B_0 \sim 1 \mu{\rm G}$ 
we find an (extremely generous) upper limit 
of $d \sim 5 \times 10^{-10}$, a value 
much smaller than the 
intrinsic polarization degree 
of CMB anisotropies and of the polarization anisotropies 
``directly'' induced by cosmic magnetic fields (see, e.g., Barrow et al. 
(1997), Subramanian et al. (2003), and references therein)
and well below any observational chance.

\section{Acknowledgements}

It is a pleasure to thank L. Danese, G. De Zotti, 
G. Giovannini, L. Moscardini, and R. Salvaterra
for useful discussions and collaborations. 
Some of the calculations presented here have been
carried out on an alpha digital unix machine
at the IFP/CNR in Milano by using some NAG integration codes.
We warmly thank an anonymous referee for a stimulating 
comment.

%\newpage
\end{document}